\documentclass[10pt, conference]{IEEEtran}
\IEEEoverridecommandlockouts

\usepackage{cite}

\usepackage{amsmath,amssymb,amsfonts}
\usepackage{algorithmic}
\usepackage{graphicx}
\usepackage{textcomp}
\usepackage{xcolor}
\usepackage[font=small]{subcaption}
\usepackage[font=small,labelfont=bf]{caption}
\usepackage{comment}
\usepackage{multirow}
\usepackage[hyphens]{url}
\usepackage{tikz}

\usepackage{titlesec}

\titlespacing*{\section}{0pt}{1ex}{1ex}  
\titlespacing*{\subsection}{0pt}{0.5ex}{0.5ex}  

\def\BibTeX{{\rm B\kern-.05em{\sc i\kern-.025em b}\kern-.08em
    T\kern-.1667em\lower.7ex\hbox{E}\kern-.125emX}}
\begin{document}

\title{Distributed Uplink Joint Transmission for 6G Communication\\
\thanks{Efforts sponsored by the U.S. Government under the Training and Readiness Accelerator II (TReX II), OTA. The U.S. Government is authorized to reproduce and distribute reprints for Governmental purposes notwithstanding any copyright notation thereon. The views and conclusions contained herein are those of the authors and should not be interpreted as necessarily representing the official policies or endorsements, either expressed or implied, of the U.S. Government.}
}
\author{\IEEEauthorblockN{
         Kumar Sai Bondada\IEEEauthorrefmark{1},
         Usama Saeed\IEEEauthorrefmark{1},
         Yibin Liang\IEEEauthorrefmark{1},
         Daniel J. Jakubisin\IEEEauthorrefmark{1}\IEEEauthorrefmark{2},
         Lingjia Liu\IEEEauthorrefmark{1},
         R. Michael Buehrer\IEEEauthorrefmark{1}
} 
     \IEEEauthorblockA{
         \IEEEauthorrefmark{1}Wireless@VT, Bradley Department of ECE, Virginia Tech, Blacksburg, VA, USA \\
         \IEEEauthorrefmark{2}Virginia Tech National Security Institute, Blacksburg, VA, USA  
         }
           \vspace{-1cm}
         }
       
\maketitle

\begin{abstract}
This paper investigates the spectral efficiency achieved through uplink joint transmission, where a serving user and the network users (UEs) collaborate by jointly transmitting to the base station (BS). The analysis incorporates the resource requirements for information sharing among UEs as a critical factor in the capacity evaluation. Furthermore, coherent and non-coherent joint transmission schemes are compared under various transmission power scenarios, providing insights into spectral and energy efficiency. A selection algorithm identifying the optimal UEs for joint transmission, achieving maximum capacity, is discussed. The results indicate that uplink joint transmission is one of the promising techniques for enabling 6G, achieving greater spectral efficiency even when accounting for the resource requirements for information sharing.
\end{abstract}

\vspace{-2mm}
\section{Introduction \& Related Works}
\begin{figure}[h]
\centerline{\includegraphics[width=0.7\linewidth]
{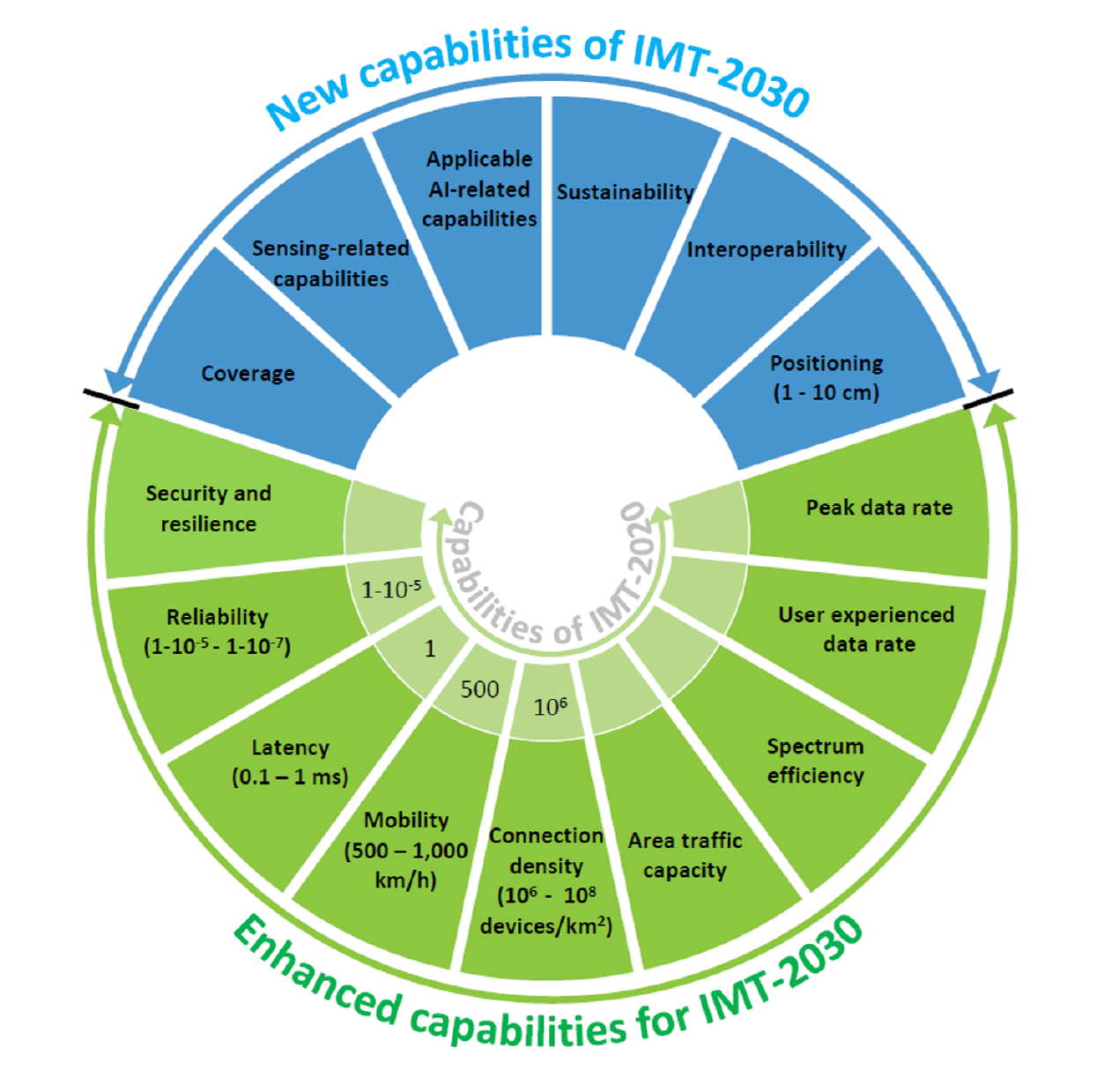}}
\caption{IMT-2030 6G Capabilities. (figure from \cite{ITU2024})}
\label{fig:6G_capabilities}
\vspace{-1.5em}
\end{figure}

6G new and extended capabilities, as shown in Fig. \ref{fig:6G_capabilities}, demonstrate highly ambitious performance goals to support usage scenarios such as ubiquitous, intelligent, and integrated sensing, along with massive, immersive, hyper-reliable, and low-latency communications. Various peak data rates, ranging from 50 Gbps to 200 Gbps, alongside user-experienced rates starting at 300 Mbps, have been proposed.  6G is designed to support a wide range of applications, including AR/VR, holographic communication, large-scale IoT, positioning, autonomous systems, healthcare, and more, providing a comprehensive solution for emerging technological demands. With new applications requiring wireless networked control of robotic systems, such as autonomous driving and smart factories, the trend is moving toward decentralized control with cooperative communications.  

\begin{figure*}[h]
\centering
\begin{subfigure}[t]{0.35\textwidth}
    \centering
    \includegraphics[scale=0.36]{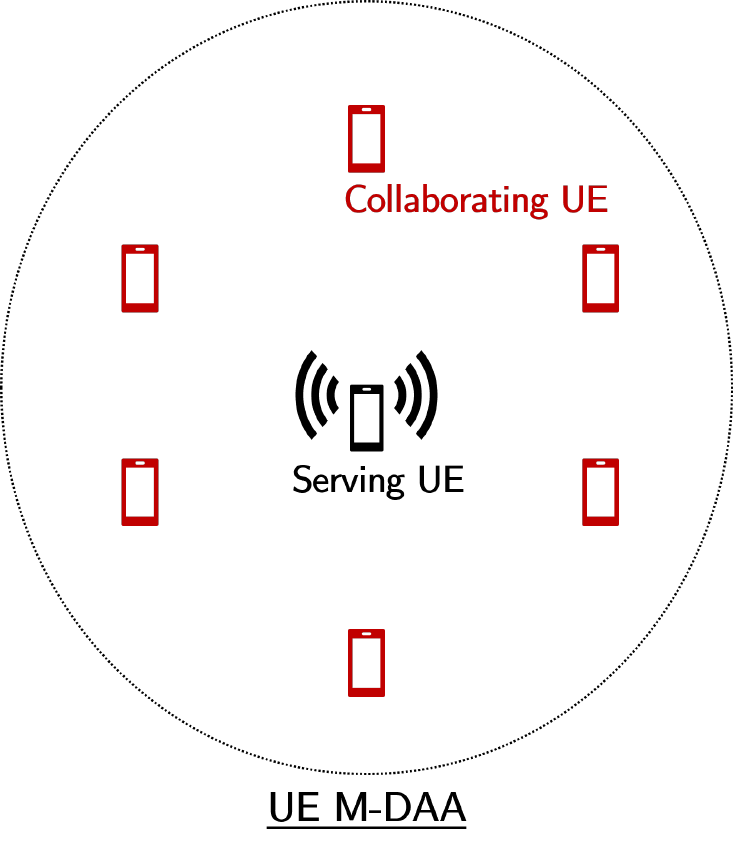}
    \caption{Phase 1: Multicasting}
    \label{subfig:phase1}
\end{subfigure}
~
\begin{subfigure}[t]{0.6\textwidth}
    \centering
    \includegraphics[scale=0.36]{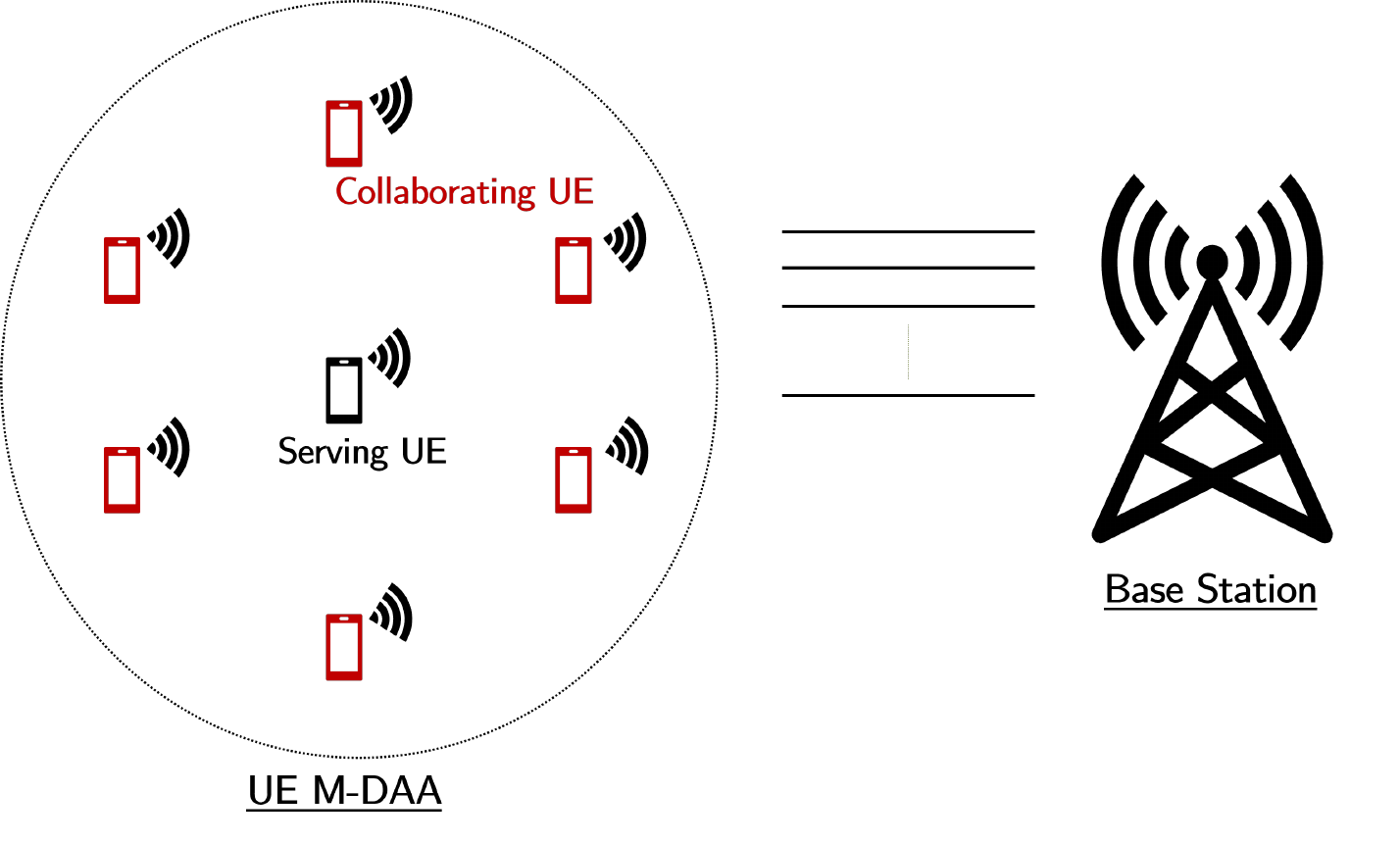}
    \caption{Phase 2: Joint transmission to the BS}
    \label{subfig:phase2}
\end{subfigure}
\caption{Illustration of the two-phase process for uplink joint transmission. In Phase 1, the serving UE multicasts data to other UEs in the network, enabling these UEs to collaborate effectively. In Phase 2, the serving UE and collaborating UEs (shown in red) form a Mobile Distributed Antenna Array (M-DAA), enabling joint transmission to the BS.}
\label{fig:distributed_uplink_joint_transmission}
\vspace{-2 mm}
\end{figure*}

We propose uplink joint transmission as one of the techniques to meet the demanding requirements of 6G. The serving user collaborates with other users (UEs) in the network, and these UEs jointly transmit to the base station (BS), thereby increasing the received signal quality and spectral efficiency. Prior research studied collaboration among UEs through various analyses, including BLER, range extension, spectral and energy efficiency, and objectives such as tactical communications and mobile ad hoc networks \cite{1040381,4202181, 6189962,10017959,6735697, 6319679}. Research in \cite{1040381} explored cooperation among the users acting as a relay to the serving user in the downlink. The trade-off between synchronization overhead and beamforming gains with the impact of the phase errors was studied in \cite{4202181}. Null-beamforming with multiple transmitters cooperating and with feedback from multiple receivers was researched in \cite{6319679}. Studies \cite{4785387, 9492307, 6804225} examine the challenges of distributed transmit beamforming, such as the availability of accurate channel state information (CSI) and synchronization among transmitters. However, prior research has focused on joint transmission with single antenna users, without accounting for the overhead of information sharing among transmitters when analyzing the system. Other distributed MIMO (D-MIMO) systems include coordination of access points (APs) of the BS to serve the UEs for mitigating inter-cell interference and providing uniform QoS \cite{6495775,7227028,6574665}.

In this paper, a distributed uplink architecture is analyzed by evaluating the achievable capacity gains and identifying the key contributing factors. The study accounts for information sharing across transmitters and introduces a user selection algorithm to achieve optimal capacity. Furthermore, the implications of precoding are examined in both the information-sharing and joint transmission phases under varying transmission power scenarios. To facilitate uplink joint transmission, relevant enhancements to existing standards are also proposed.

The remainder of this paper is structured as follows: Section II presents the system architecture. Section III discusses the proposed additions to standards for enabling uplink joint transmission. Section IV provides the simulation results, and Section V concludes the paper with a discussion of future research directions.

\vspace{-1.5 mm}
\section{System Architecture}
We consider a system architecture as shown in Fig.~\ref{fig:distributed_uplink_joint_transmission}, comprising serving user equipment (UE) and collaborating UEs, which together form a mobile distributed antenna array (M-DAA) that jointly transmits to the BS. The collaborating UEs refer to other users in the network. Joint transmission is classified into coherent joint transmission (CJT) and non-coherent joint transmission (NCJT). Both transmission schemes require additional resources, making it necessary for the serving UE's information to be available to all collaborating UEs. To achieve this, the serving UE wirelessly multicasts its information to these UEs. Furthermore, CJT requires accurate CSI available to all UEs and strict synchronization in terms of time, frequency, and phase among the UEs.

The transmission is divided into two phases. The first phase involves serving UE multicasting its data to the collaborating UEs with bandwidth $B_1$. The second phase involves joint transmission from the UEs of M-DAA to the BS with bandwidth $B_2$. Since all collaborating UEs receive the same data, the Phase 1 capacity is determined by the UE with the lowest achievable rate. Consequently, this UE limits the Phase 2 capacity, as the Phase 2 transmission rate cannot exceed the information successfully decoded in Phase 1. 

\subsection{Phase 1 multicasting}
Let $N_t^{\text{UE}}$ and $N_r^{\text{UE}}$ represent the number of transmit and receive antennas, respectively, for each UE. The received signal at the $i$-th UE is given by
\footnote{Matrices are denoted by boldface uppercase letters, vectors by boldface lowercase letters, and italic lowercase letters are used for scalars. $|\cdot|$ indicates the determinant of the matrix.}
\[
    \mathbf{y}^{\text{UE}}_i = 
    \sqrt{G_{i}^{\text{UE}}} \mathbf{H}_{i}^{\text{UE}} \sqrt{\frac{E^{\text{UE}} }{N_t^{\text{UE}}}} 
    \mathbf{F}^{\text{UE}} \mathbf{s}^{\text{UE}} + \mathbf{n}^{\text{UE}}_i,
\]
where $G_{i}^{\text{UE}}$ represents the large-scale path loss gain and $\mathbf{H}_{i}^{\text{UE}}$ (of size $N_r^{\text{UE}} \times N_t^{\text{UE}}$) denotes the channel matrix, containing the small-scale independent complex Gaussian distributed coefficients between the serving UE and the $i$-th collaborating UE. The vector $\mathbf{s}^{\text{UE}}$ represents the data streams (of size $N_s^{\text{UE}} \times 1$) multicasted, and $\mathbf{n}^{\text{UE}}_i \sim \mathcal{N}(0, \sigma_{\text{UE}}^2 \mathbf{I}_{N_r^{\text{UE}}})$ is the noise vector per UE. The rate for the $i$-th collaborating UE is given as
\begin{equation*} 
    R_{i}^{\text{UE}} = \frac{C_{i}^{\text{UE}}}{B_1} = \log_2 \left| \mathbf{I}_{N_r^{\text{UE}}} + \frac{E^{\text{UE}} }{N_t^{\text{UE}} } \frac{\mathbf{\bar{H}}_{i}^{\text{UE}} \mathbf{F}^{\text{UE}} \left(\mathbf{F}^{\text{UE}}\right)^\text{H} \left(\mathbf{\bar{H}}_{i}^{\text{UE}}\right)^\text{H}}{\sigma_{\text{UE}}^2} \right|    
\end{equation*}
where $\bar{\mathbf{H}}_{i}^{\text{UE}} = \sqrt{G_i^{\text{UE}}} \mathbf{H}_i^{\text{UE}}$ and $E^{\text{UE}} = P^{\text{UE}} \cdot T$ is the energy per transmitted symbol, with $P^{\text{UE}}$ representing the transmit power and $T$ the symbol duration. 


The common precoder $\mathbf{F}^{\text{UE}}$ (of size $N_t^{\text{UE}} \times N_s^{\text{UE}}$) can be designed to maximize the minimum rate, ensuring that the UE with the minimum rate does not limit the overall capacity. This precoder can be solved numerically by formulating as a convex optimization problem
\begin{align*}
    \max_{\| \mathbf{F}^{\text{UE}} \|_F^2 \le N_t^{\text{UE}}} \min_{i} R_i^{\text{UE}}.
\end{align*}
This becomes a non-convex problem~\cite{6364357} if $N_s^{\text{UE}} < N_t^{\text{UE}}$. Prior research demonstrated approaches to solve the precoder using Semi-Definite Programming or Second-Order Cone Programming by relaxing the rank constraint and employing approximation techniques~\cite{article_Bengtsson,liu2024surveyrecentadvancesoptimization}. We consider $N_s^{\text{UE}} = N_t^{\text{UE}}$, allowing full rank to transmit the maximum number of independent streams, removing the rank constraint. Alternatively, an identity precoder not requiring CSI between the UE and M-DAA UEs can be used, as it is straightforward to implement. The capacity of Phase 1 with no CSI is expressed as
\begin{align*}
    R_1= \frac{C_1}{B_1} = \min \left( R_{i}^{\text{UE}} \right).
\end{align*}

\subsection{Phase 2 Coherent Joint Transmission}
In phase 2, the serving UE, along with $U-1$ collaborating UEs, jointly transmits to the BS, as shown in Fig.~\ref{subfig:phase2}. Considering $\mathbf{H}_i^{\text{BS}}$ as the channel matrix filled with independent complex Gaussian distributed coefficients and the large-scale path loss gain $G_i^{\text{BS}}$ between the $i$-th UE and the BS, the received signal at the BS is given by
\begin{align}
\mathbf{y}^{\text{BS}} = \sum_{i=1}^{U} \sqrt{G^{\text{BS}}_i} \mathbf{H}_i^{\text{BS}} \sqrt{\frac{E^{\text{UE}}}{N_t^{\text{UE}}}} \mathbf{F}_i \mathbf{s} + \mathbf{n}^{\text{BS}},  
\label{eq:phase2_1}
\end{align}
which can be rewritten in a compact matrix form as
$
\mathbf{y}^{\text{BS}} =\sqrt{\frac{E^{\text{UE}}}{N_t^{\text{UE}}}} \mathbf{H} \mathbf{F} \mathbf{s} + \mathbf{n}^{\text{BS}},    
$
where
\[
\mathbf{H} = \begin{bmatrix} \sqrt{G^{\text{BS}}_1} \mathbf{H}_1^{\text{BS}} & \sqrt{G^{\text{BS}}_2} \mathbf{H}_2^{\text{BS}} & \dots & \sqrt{G^{\text{BS}}_U} \mathbf{H}_U^{\text{BS}} \end{bmatrix}_{N^{\text{BS}}_r \times U . N^{\text{UE}}_t}
\]
is the effective channel matrix ($N^{\text{BS}}_r$ is the number of BS receive antennas), formed by horizontal stacking each UE's channel matrix, and
$
\mathbf{F} =   \begin{bmatrix}
\mathbf{F}_1 &
\mathbf{F}_2 &
\hdots & 
\mathbf{F}_U
\end{bmatrix}^{^\text{T}}
$
is the aggregate precoder formed by vertical stacking each UE precoder.
Each UE's precoder $\mathbf{F}_i$ is of size $N_t^{\text{UE}} \times \bar{N}_s$ and $\bar{N}_s = \min(U \cdot N_t^{\text{UE}}, N^{\text{BS}}_r)$ corresponds to the number of independent data streams. The common data vector $\mathbf{s}$ is of size $\bar{N}_s \times 1$, and the noise vector $\mathbf{n}^{\text{BS}} \sim \mathcal{N}(0, \sigma_{\text{BS}}^2 \mathbf{I}_{N^{\text{BS}}_r})$. The precoder $\mathbf{F}$ is constructed based on the composite channel $\mathbf{H}$ through singular value decomposition (SVD), i.e., 
$\mathbf{H} = \mathbf{U} \mathbf{\Sigma} \mathbf{V}^\text{H}$. The precoder $\mathbf{F}$ is obtained from the right singular vectors corresponding to the non-zero singular values of $\mathbf{H}$, i.e., $\mathbf{V}_{(:, 1:\bar{N}_s)}$, multiplied by a diagonal matrix $\mathbf{P}$ of size $\bar{N}_s$. The diagonal elements of $\mathbf{P}$ represent the power allocated to each stream, satisfying $\sum_{i=1}^{\bar{N}_s}P_{i}=U \cdot N_t^{\text{UE}}$. This constraint ensures UEs' individual power constraints are met. The Phase 2 capacity is given by
\begin{align*}
     R_2^{\text{CJT}} = \frac{C_2^{\text{CJT}}}{B_2}  = \sum_{i=1}^{\bar{N}_s}\log_2\left(1 +  \frac{E^{\text{UE}}}{N_t^{\text{UE}} \sigma^2_{\text{BS}}} \lambda_i(\mathbf{H}^\text{H}\mathbf{H})  \mathbf{P}_i \right),
\end{align*}
where $\lambda_i(\mathbf{H}^\text{H}\mathbf{H})$ are the eigenvalues of $\mathbf{H}^\text{H}\mathbf{H}$, and $\mathbf{P}_i$ is the capacity-achieving power allocation per stream given by the water-filling algorithm \cite{Heath}.

\subsection{Phase 2 Non-Coherent Joint Transmission}
CJT involves strict synchronization requirements and accurate CSI availability among the UEs. It also requires the channel coherence time to be large so that the channel remains stable and the precoder is aligned with the channel. We now consider NCJT, where the synchronization requirements are relaxed and the precoder is not needed. NCJT is useful when the channels vary rapidly, i.e., in high-mobility scenarios. The system considers the number of BS receive antennas greater than the total number of UE transmit antennas. The BS can decode up to $N_r^{\text{BS}}$ streams, provided that the UEs transmit at least that number of streams. Considering $N_t^{\text{UE}} = 2$ and $N_r^{\text{BS}} = 4$, the users can be grouped into two groups $\{u_i\}$ and $\{u_j\}$, where $i=1,...,N_1$, $j=1,..,N_2$, and $N_1 + N_2=U$. Each group transmits separate streams, in total of 4 different streams for the BS. The received signal at the BS is given as
\begin{multline*}
\mathbf{y}^{\text{BS}} =  \sqrt{\frac{E^{\text{UE}}}{N_t^{\text{UE}}}} \left( \sum_{i} \sqrt{G_i^{\text{BS}}} \mathbf{H}_i^{\text{BS}} \mathbf{s}{'} +  \sum_{j} \sqrt{G_j^{\text{BS}}} \mathbf{H}_j^{\text{BS}} \mathbf{s}{''}\right) \\ + \mathbf{n}^{\text{BS}},    
\end{multline*}
where $s'$ and $s''$ are the transmitted streams from different groups. The expression can be simplified to 
\begin{equation*}
\mathbf{y}^{\text{BS}} =  \sqrt{\frac{E^{\text{UE}}}{N_t^{\text{UE}}}}  \mathbf{H}^{\text{NCJT}} 
\begin{bmatrix} \mathbf{s}' \\ \mathbf{s}'' \end{bmatrix}   + \mathbf{n}^{\text{BS}},        
\end{equation*}
where $\mathbf{H}^{\text{NCJT}} = \begin{bmatrix} \mathbf{H}' & \mathbf{H}'' \end{bmatrix}$, $\mathbf{H}' = \sum_{i} \sqrt{G_i^{\text{BS}}} \mathbf{H}_i^{\text{BS}}$ and $\mathbf{H}'' = \sum_{j} \sqrt{G_j^{\text{BS}}} \mathbf{H}_j^{\text{BS}} $. The rate can be expressed as 
\begin{align*}
R_2^{\text{NCJT}} = \frac{C_2^{\text{NCJT}}}{B_2} = \log_2\bigg| \mathbf{I}_{N_r^{\text{BS}}} + \frac{E^{\text{UE}}}{N_t^{\text{UE}}}  \frac{\mathbf{H}^{\text{NCJT}}  \big(\mathbf{H}^{\text{NCJT}}\big)^{\text{H}}  }{\sigma^2_{\text{BS}}}   \bigg|.
\end{align*}

\section{Standardization Efforts Enabling Uplink Joint Transmission}
\subsection{Sidelink} 
Sidelink enables users to communicate directly, with or without the involvement of a cellular network. The 3rd Generation Partnership Project (3GPP) Technical Report (TR) 21.918 \cite{3gpp_sidelink_sTR} introduced several new work items in Release 18 to enhance sidelink capabilities, including relay enhancements. These relay improvements facilitate direct communication between UE devices or between UEs and the network through multipath relay, where in-coverage UEs act as gateways to the BS. Prior research \cite{9392777} also highlights unicast, multicast, and groupcast communication through sidelink. These advancements motivate the adoption of D-MIMO in uplink. Building upon these developments, the proposed feature leverages networked UEs as collaborating UEs (as demonstrated in our prior work, where UEs act as APs for the BS \cite{10757642}), promising significant capacity gains in alignment with 6G objectives.

\subsection{Synchronization}
Synchronization is classified into two types. 

\subsubsection{Intra M-DAA Synchronization}
It involves the collaborating UEs synchronizing with the serving UE in time, frequency, and phase. Due to external factors such as aging, temperature, and voltage fluctuations, the UE (oscillator) oscillates differently, causing deviations in time references and carrier frequencies from their intended values. These asynchronous oscillators hinder the realization of the potential gains achievable through CJT. Various techniques have been developed to estimate and correct the effects of oscillator asynchrony, either digitally by incorporating correction parameters into the baseband signal or by directly controlling the oscillator using an external reference signal. Prior research \cite{7218555,6624252,8742232,3448623,9994246,8396683} has demonstrated picosecond-level time accuracy and frequency accuracy within 1–2 Hz under various conditions. While GPS could be used, it is a power-intensive solution that significantly drains UE batteries. 

\subsubsection{Inter M-DAA to BS Synchronization}
It involves synchronization between the UE M-DAA and the BS, encompassing uplink and downlink. In the uplink, the UE M-DAA determines the precise timing for initiating CJT. In the downlink, UEs detect the radio frame and the OFDM symbol boundary. Standard 3GPP synchronization procedures can be applied. For the downlink, the Synchronization Signal (SS) Block is typically used, while for the uplink, PRACH is employed.

\subsection{Frame Structure}
3GPP TR 38.785 prioritizes TDD for sidelink operation \cite{tr38785}. 
The time slots reserved for two-way communication between UEs in sidelink could be divided into time slots reserved for Phase 1 and Phase 2 of the D-MIMO architectures outlined in this work.
Further, TR 38.785 also defines sidelink operation to occupy the Uplink band, which ties in well with the uplink D-MIMO scenario.
However, additional intra-UE signaling would be needed to enable D-MIMO operation, such as for Downlink (gNB to UE) CSI between UEs, communicating UE selection for CJT, and cluster formation for NCJT. The synchronization signals for performing intra-UE M-DAA synchronization should be added.

\begin{table}[h]
    \centering
    \caption{Simulation Parameters}
    \begin{tabular}{|c|c|}
        \hline
        \textbf{Parameter} & \textbf{Value} \\
        \hline
        Precoding & SVD Beamforming \\
        BS \& UE heights & 20 \& 2 meters \\   
        Pathloss model & UMi \\
        ($N^t_{\text{BS}}$,$N^t_{\text{UE}}$,$N^r_{\text{UE}}$)  & (4,2,2) \\ 
        Carrier frequency & 7.5 GHz\\ 
        Bandwidth ($B_1, B_2$) & 10 Mhz \\ 
        User Phase 1 \& Phase 2 Tx power & 26 \& 23 dBm \\   
        Avg. height of buildings & 20 meters \\
        Noise Figure of BS \& UE & 9 \& 4 \\
        \hline
    \end{tabular}
    \label{tab:sim}
\end{table}

\vspace{-1em}
\section{Simulation Results \& Insights}
We consider an urban micro (UMi) path-loss setting with simulation parameters listed in Table~\ref{tab:sim}. The UE's phase 1 transmission power is set to 26 dBm. For phase 2, each UE's transmission is considered identical. The maximum transmission power supported by each UE for uplink is 23 dBm. The analysis is conducted in two stages. First, the individual phase capacities are evaluated independently. Second, the two phases are combined to study the overall system capacity, which is D-MIMO capacity. The 7.5 GHz band is considered for simulations as it is expected to be used in initial 6G roll-outs, balancing spectrum availability, coverage, infrastructure costs, and global harmonization \cite{dropmann2024golden}.

\begin{figure}[h]
    \centering
    \includegraphics[width=0.7\columnwidth]{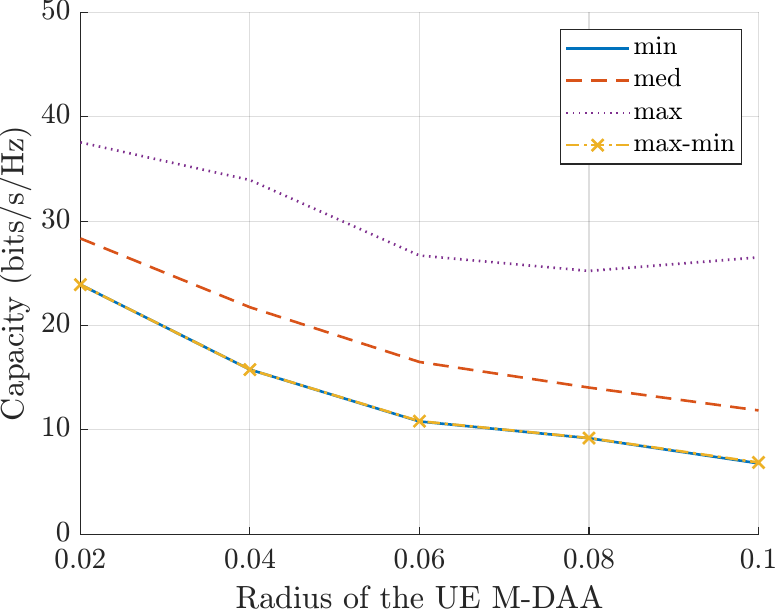}
    \caption{Average minimum, median, maximum rate of 10 collaborating UEs, and rate using max-min rate precoder in Phase 1}
    \label{fig:UE_MDAA_CAP_10_nodes_7Ghz_phase1}
\vspace{-2.5mm}
\end{figure}

 Fig.~\ref{fig:UE_MDAA_CAP_10_nodes_7Ghz_phase1} illustrates the Phase 1 average capacity, depicting the minimum, median, and maximum rates and the rate achieved with the max-min precoder, for 10 collaborating UEs. The UEs are uniformly distributed within a circular radius $R$ of M-DAA are selected. The plot indicates the capacity achieved using the max-min precoder is comparable to the minimum rate, suggesting that the precoder may not provide significant gains when the number of UEs significantly exceeds the number of antennas per UE, as the UEs have fewer degrees of freedom.

\begin{figure}[h]
  \centering
  \begin{subfigure}{\columnwidth}
  \centering
    \includegraphics[width=0.65\linewidth]{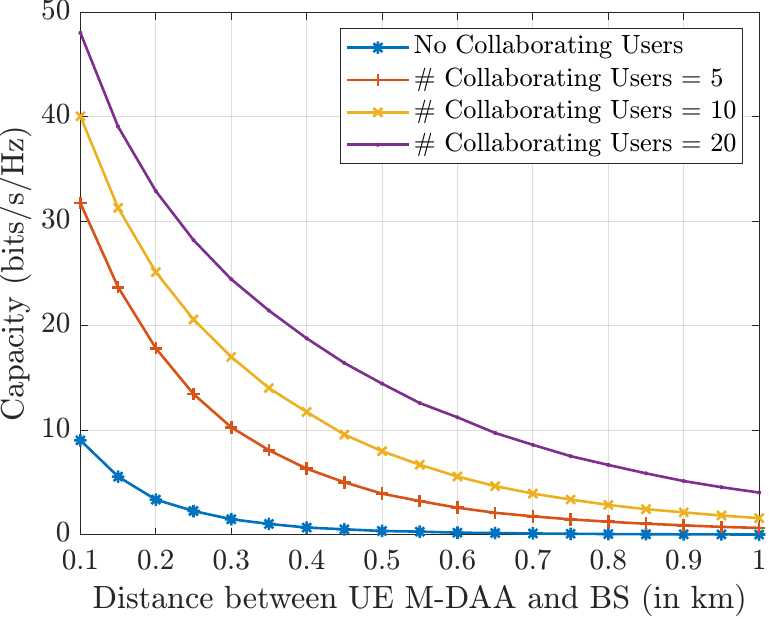}
    \caption{Phase 2 Capacity plots}
    \label{fig:phase2_cap_svd_multiple_nodes_cases}
  \end{subfigure}
  \begin{subfigure}{\columnwidth}
  \centering
    \includegraphics[width=0.65\linewidth]{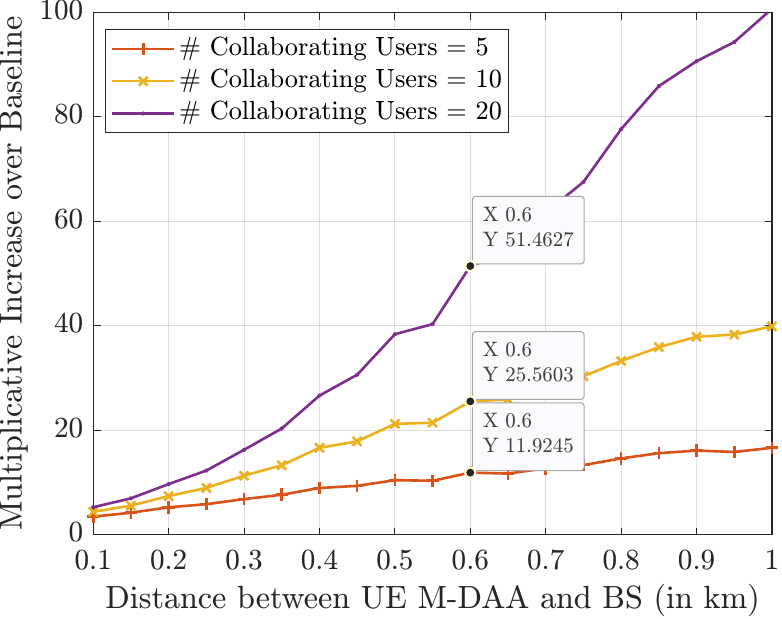}
    \caption{Phase 2 Capacity Relative improvement over the baseline}
    \label{fig:phase2_cap_svd_multiple_nodes_cases_relative}
  \end{subfigure}
    \caption{Phase 2 average capacity with full transmission power.}
  \label{fig:phase2_cap_svd_multiple_nodes_cases_full_power}
  \vspace{-3mm}
\end{figure}

\begin{figure}[h]
  \centering
  \begin{subfigure}{\columnwidth}
  \centering
    \includegraphics[width=0.65\linewidth]{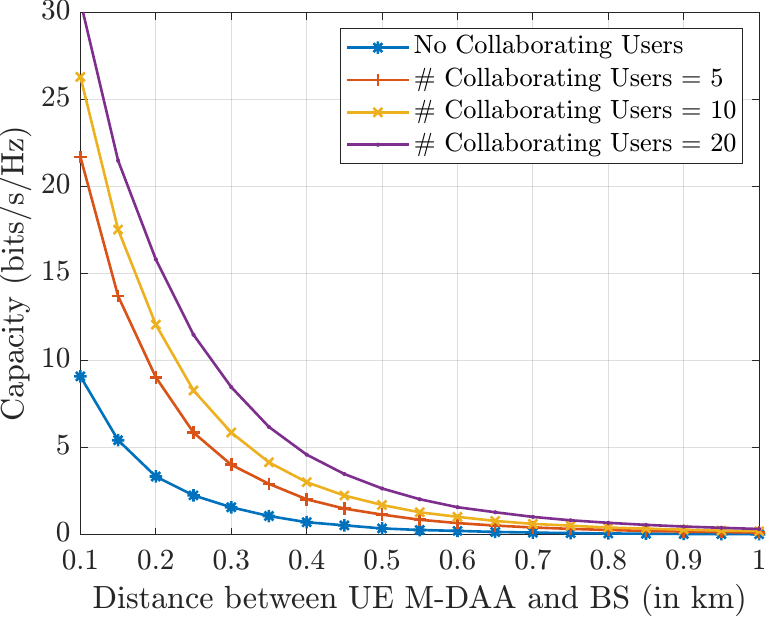}
    \caption{Phase 2 Capacity plots}
\label{fig:phase2_cap_svd_multiple_nodes_cases_normalized}
  \end{subfigure}
  \begin{subfigure}{\columnwidth}
  \centering
    \includegraphics[width=0.65\linewidth]{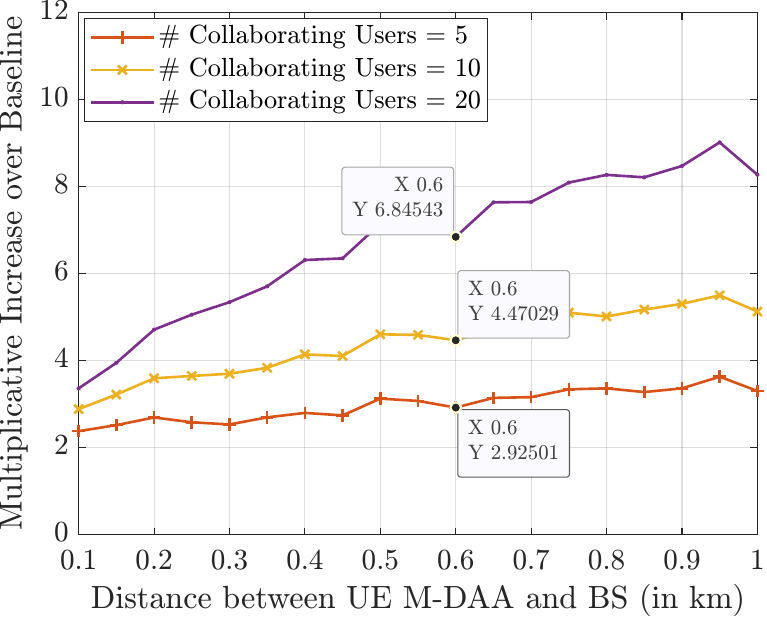}
    \caption{Phase 2 Capacity Relative improvement over the baseline}\label{fig:phase2_cap_svd_multiple_nodes_cases_normalized_relative}
  \end{subfigure}
  \caption{Phase 2 average capacity with transmission power normalized.}  \label{fig:phase2_cap_svd_multiple_nodes_cases_normalized_power}
  \vspace{-1.5em}
\end{figure}

Fig.~\ref{fig:phase2_cap_svd_multiple_nodes_cases_full_power} depicts the Phase 2 capacity (CJT) for varying numbers of collaborating UEs, each transmitting at the full transmission power of 23~dBm. Fig.~\ref{fig:phase2_cap_svd_multiple_nodes_cases} presents the Phase 2 capacity, while Fig.~\ref{fig:phase2_cap_svd_multiple_nodes_cases_relative} illustrates the relative improvements (multiplicative increase) over the baseline, where the serving UE directly communicates with the BS. The results demonstrate a significant linear relative improvement in capacity compared to the baseline, attributed to the combined effect of additional power and antennas from collaborating UEs. Fig.~\ref{fig:phase2_cap_svd_multiple_nodes_cases_normalized_power} examines the Phase 2 capacity when the transmission power of each UE is normalized such that the total transmission power of all UEs equals 23~dBm. Fig.~\ref{fig:phase2_cap_svd_multiple_nodes_cases_normalized} shows the Phase 2 capacity, and Fig.~\ref{fig:phase2_cap_svd_multiple_nodes_cases_normalized_relative} presents the relative improvements over the baseline. Although the capacity improvements are higher than the baseline, ranging from 2x to 9x, they remain smaller than the full transmission power scenario, with the gains driven solely by the additional antennas. All simulations consider a 50-meter radius for the M-DAA.

For D-MIMO capacity, the analysis considers time durations $T_1$ for Phase 1 and $T_2$ for Phase 2, with the total duration constrained by $T_1 + T_2$ = $1$ second. The capacity is determined by evaluating all possible combinations of UEs selected in Phase 1, ranging from one to ten UEs, resulting in 1024 combinations to identify the maximum achievable capacity. The number of bits transferred in 1 second is given by $C_1 \times T_1$ = $C_2 \times T_2$. For baseline comparisons, the baseline capacity is provided by $C_B \times (T_1 + T_2)$. 

\begin{figure}[h]
  \centering
    \includegraphics[width=0.7\linewidth]{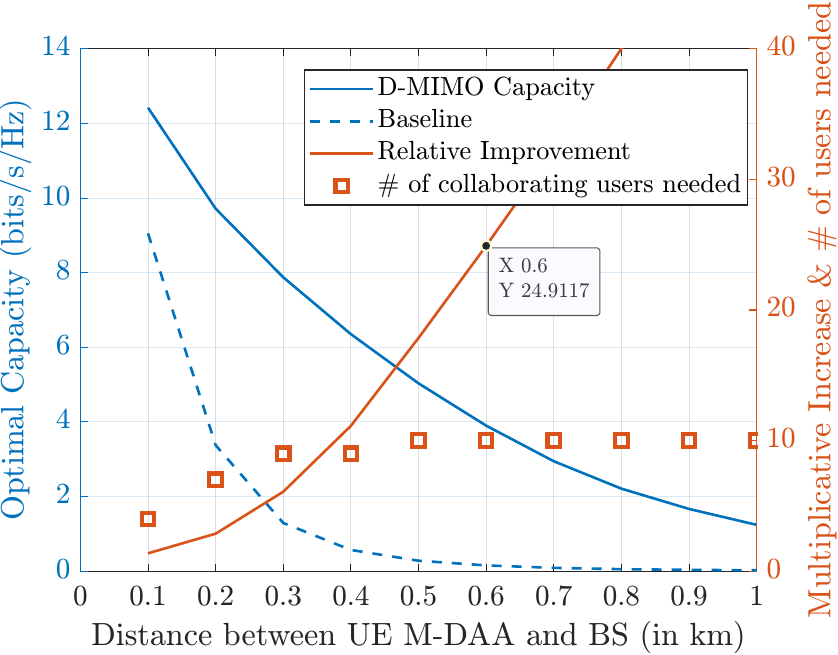}
    \caption{CJT D-MIMO average capacity with full power transmission.}
  \label{fig:optimal_cap_svd_10_full_TxP}
\end{figure}

\begin{figure}[h]
    \centering
    \includegraphics[width=0.7\linewidth]{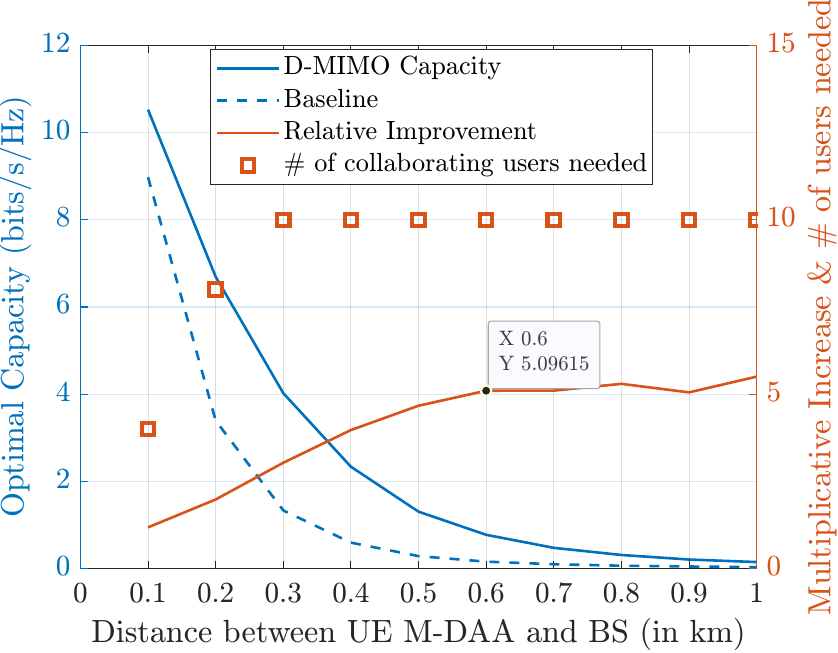}
    \caption{CJT D-MIMO average capacity with Power normalization.}
  \label{fig:optimal_cap_svd_10}
    \vspace{-2mm}
\end{figure}

Fig.~\ref{fig:optimal_cap_svd_10_full_TxP} illustrates the optimal CJT D-MIMO capacity and baseline capacity on the left side of the y-axis, the relative improvement over the baseline and the number of UEs required to achieve optimal capacity from the 10 collaborating UEs with full power transmission on the right side of the y-axis. The relative improvement and the number of UEs selected to achieve optimal capacity are dimensionless. At shorter distances, not all UEs are needed to achieve optimal capacity. However, at longer distances, all UEs become necessary, as the Phase 2 capacity weakens with fewer UEs, requiring the participation of all UEs. Additionally, Fig.~\ref{fig:optimal_cap_svd_10} illustrates the D-MIMO capacity under normalized power transmission conditions. Since power is normalized, more UEs are required even at shorter distances to achieve optimal capacity. Comparing the relative improvements, they are significantly higher for full power transmission than in the normalized case.

\begin{figure}[h]
    \centering
    \includegraphics[width=0.7\linewidth]{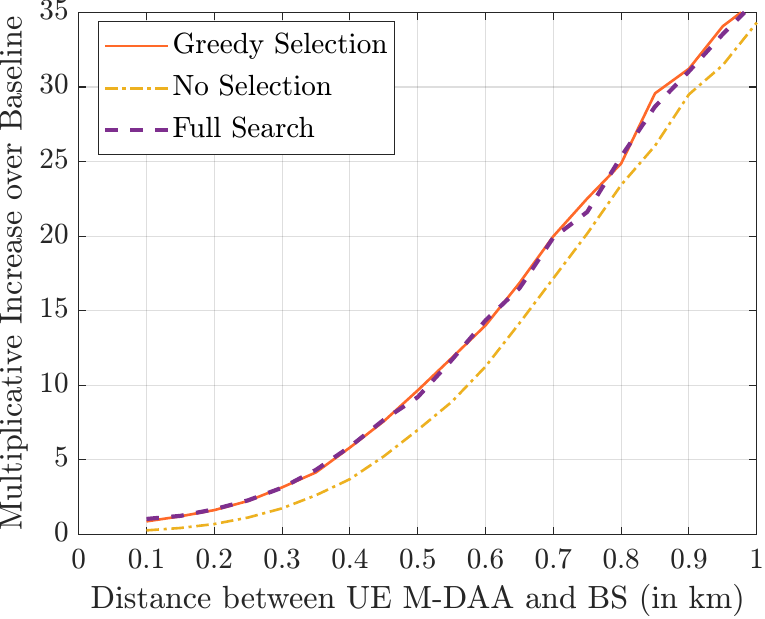}
    \caption{Capacity improvements using different selection algorithms.}
  \label{fig:optimal_cap_svd_10_selection-algorithm}
    \vspace{-2mm}
\end{figure}

In order to develop a user selection algorithm, it is first important to identify the bottlenecks of the system.
We note that the addition of UEs in CJT Phase 2 adds transmit power in addition to transmit diversity. 
However, the addition of a low-capacity link in Phase 1 may cause error propagation in Phase 2. Therefore, we introduce a greedy algorithm for user selection that prioritizes Phase 1 (treating it as the bottleneck) but takes into account information from both phases.
We start by sorting UEs in descending order based on Phase 1 capacities $R_i^{\text{UE}}$. We then select an initial subset $\mathcal{S}_K$ of the top $K$ UEs, such that $R_K^{\text{UE}}\geq R_{2,\mathcal{S}}^{\text{CJT}}$, where $R_{2,\mathcal{S}}^{\text{CJT}}$ is the capacity of CJT Phase 2 provided the UEs in set $\mathcal{S}$ are selected. We may then iteratively add one UE at a time to $\mathcal{S}$ and stop when adding a UE to $\mathcal{S}$ reduces the total system capacity $\min(R_1, R_2^{\text{CJT}})$. Fig.~\ref{fig:optimal_cap_svd_10_selection-algorithm} shows the relative capacity over the baseline with the user selection algorithm, with all the UEs selected, and full search in UE M-DAA radius 200 meters. The user selection algorithm meets the optimal capacity.

 \vspace{-1em}

  \vspace{-2em}

 \vspace{-1em}
 
\begin{figure}[h]
  \centering
    \includegraphics[width=0.7\linewidth]{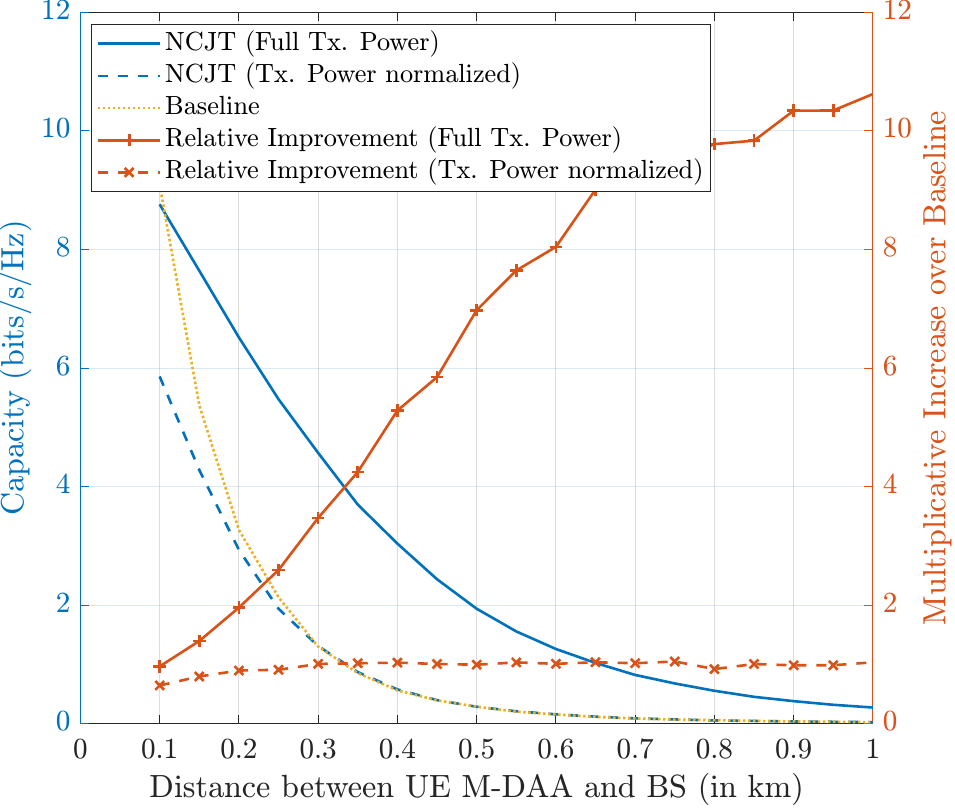}
  \caption{NCJT D-MIMO average capacity.}
  \label{fig:optimal_cap_svd_10_node_NCJT}
  \vspace{-2em}
\end{figure}

Fig.~\ref{fig:optimal_cap_svd_10_node_NCJT} shows the optimal NCJT D-MIMO capacity performance plots. The improvements compared to CJT are minimal. In the second phase, the number of layers considered is four, with one cluster including the serving UE transmitting the first two streams and the second cluster transmitting the next two streams. The number of UEs in each cluster is equally divided. NCJT normalized power transmission has no benefits compared to the baseline capacity.

\section{Conclusion}
Despite the overhead in Phase 1, the D-MIMO architecture proposed with CJT achieves significant capacity gains under full transmission power and moderate improvements with power normalization. Similarly, D-MIMO with NCJT offers moderate capacity enhancements with full transmission power, though power normalization yields no added advantage. Phase 1 is constrained by the user with the minimum rate and the use of a precoder in Phase 1 shows no significant benefit. Additionally, D-MIMO with CJT faces strict synchronization requirements, with challenges arising from CSI availability and synchronization offsets. These findings underscore the role of D-MIMO as one of the key techniques enabling 6G, with proposed standardization efforts aiming to incorporate synchronization and information sharing among users for practical deployment. The impact of synchronization offsets and mobility on the performance of CJT and NCJT should be understood as the next step. If the performance is not satisfactory, the focus should be on improving the synchronization algorithms, and the trade-off to switch between CJT and NCJT depending on the mobility conditions to be explored.



\bibliographystyle{IEEEtran}
\bibliography{reference}

\end{document}